\documentclass{iaus}
\usepackage{graphicx}

\title[Large-scale nebula around the ULX IC342 X-1] 
{Large-scale radio nebula around the Ultra-Luminous X-ray Source IC 342 X-1}

\author[D. Cseh, C. Lang, S. Corbel, P. Kaaret, F. Gris\'e]   
{D\'avid Cseh$^1$, Cornelia Lang$^2$, St\'ephane Corbel$^1$, Philip Kaaret$^2$, Fabien Gris\'e$^2$}

\affiliation{$^1$Laboratoire Astrophysique des Interactions Multi-echelles (UMR 7158),\\
\,\,\,CEA/DSM-CNRS-Universite Paris Diderot, CEA Saclay, F-91191 Gif sur Yvette, France\\ email: {\tt david.cseh@cea.fr} \\[\affilskip]
$^2$Department of Physics and Astronomy, University of Iowa, Iowa City, IA 52242, US\\}

\pubyear{2010}
\volume{275}  
\pagerange{1--2}
\jname{Jets at all Scales}
\editors{Gustavo E. Romero, Rashid A. Sunyaev and Tomaso Belloni, eds.}
\begin{document}

\maketitle

\begin{abstract}
We present discovery of a radio nebula associated with the ultraluminous X-ray source (ULX) IC 342 X-1 using the Very Large Array (VLA). Taking the surrounding nebula as a calorimeter, one can constrain the intrinsic power of the ULX source.  We compare the obtained power that is needed to supply the radio nebula with the W50 nebula powered by the microquasar SS433 and with other ULXs. We find that the power required is at least two orders of magnitude greater than that needed to power radio emission from the W50 nebula associated with the microquasar SS433. In addition, we report the detection of a compact radio core at the location of the X-ray source.

\end{abstract}

ULXs are variable off-nuclear X-ray sources in external galaxies with luminosities greatly exceeding the Eddington luminosity of a stellar-mass compact object, assuming isotropic emission (\cite{CM}). The irregular variability, observed on time scales from seconds to years, suggests that ULXs are binary systems containing a compact object that is either a stellar-mass black hole with beamed (\cite{king,elmar}) or super-Eddington emission (\cite{beg}) or an intermediate mass black hole.

We present the discovery of an associated radio nebula around IC 342 X-1 using the VLA. We detected the source at a 10-$\sigma$ level with peak intensity of $\sim$120 $\mu$Jy (Fig.1., left) and the estimated flux density is $\sim2$~mJy. We estimated the total energy budget of the radio nebula assuming radiation via synchrotron emission, equipartition between particles and fields, and equal energy in electrons and baryons. A radio spectral index of -0.8 of NGC 5408 X-1 was assumed and we use a lower frequency cutoff of 1.3 GHz and an upper frequency cutoff of 6.2 GHz. For a source diameter of $\sim$220 pc and a filling factor of unity, we find that the total energy required to power the radio nebula is $9 \times 10^{50}$ erg, (with an equipartition magnetic field of 7 $\mu$G). 

Only a handful of radio detections of ULXs have been made so far (\cite{phil,miller,soria,cornelia}). These sources all show large nebulae ($>$35 pc) that are likely powered by continuous energy input from the ULX, in the same manner as the W50 nebula is powered by the Galactic binary SS 433 (\cite{dubner}). However, the ULX radio nebulae require very high total energy content,  10$^{49}$ erg as compared with 10$^{46}$ erg for W50. {\it The energy of the nebula around IC342 X-1 is similar to ULX NGC5408 X-1 and Holmberg II X-1 but at least 2 orders larger than SS433.}

We detected a radio point source at the location of IC342 X-1 (Fig.1, right) with a peak intensity of $\sim90 \, \mu$Jy at a 7-$\sigma$ level, which can be consistent with a compact jet (though we have no spectral information). The radio image was weighted (robust=-2) in order to resolve the diffuse nebular emission. Using the fundamental plane of black holes (eg. \cite{elmar2}) -  which is a relationship between X-ray luminosity, radio luminosity and black hole mass - we can estimate the mass of the ULX. The application of the fundamental plane requires radiatively inefficiently accreting sources, ie. hard state objects. IC 342 X-1 is one of relatively few ULXs usually found in the X-ray hard state (\cite{feng}). Substituting the flux of the point source (and $L_{X}=6 \times 10^{39}$ erg/s), we obtain for the mass of the black hole $M_{\rm{BH}}\simeq10^{4}$~M$_{\odot}$, which should be taken as an order of magnitude under the hypotheses mentioned before (Cseh \textit{et al.} in prep).

\begin{figure*}
\label{ICFOV}
\includegraphics[width=2in, angle=270]{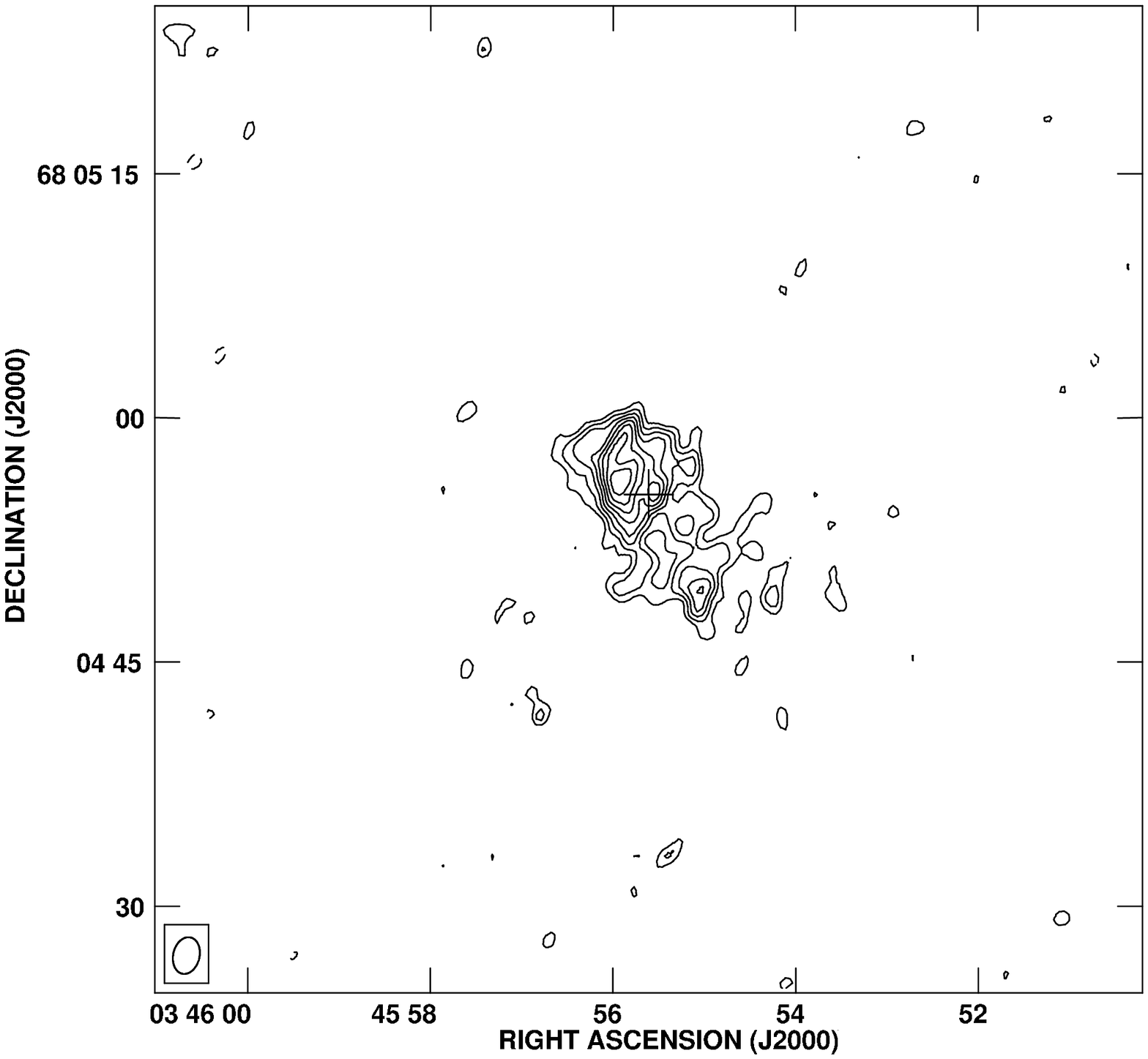}
\includegraphics[width=2.3in, angle=270]{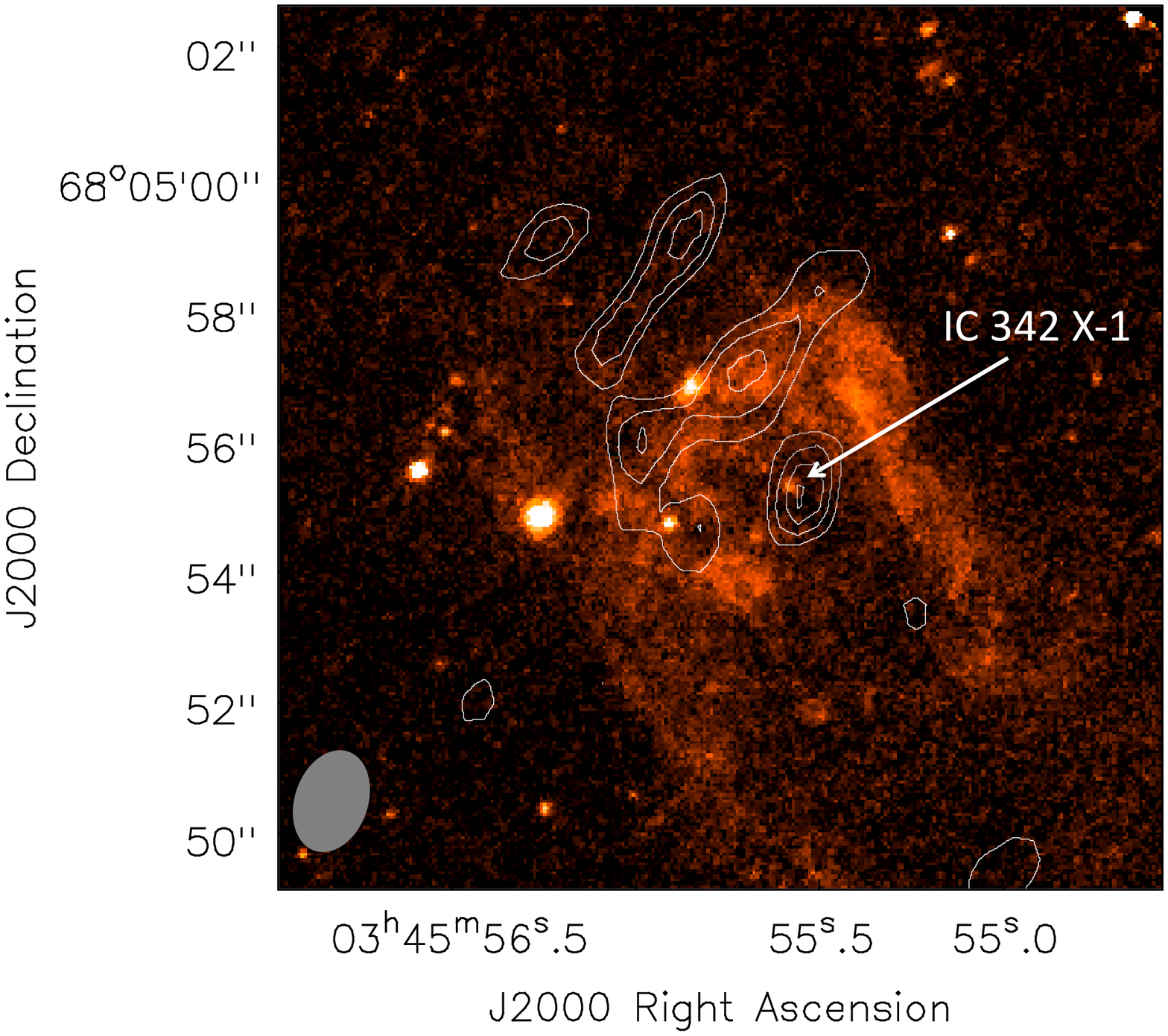}
\caption{{\bf Left:} The 5-GHz VLA B- and C-array combined image of IC 342 X-1. The first contours are drawn at 3~$\sigma$, at $\pm$~$33$~$\mu$Jy/beam. The positive contour levels increase by a factor of 1~$\sigma$. The peak brightness is 122.4 $\mu$Jy/beam. The Gaussian restoring beam is $2.3" \times 1.6"$ at PA=$-13^\circ$. The cross marks the Chandra X-ray position of the ULX source. {\bf Right:} The weighted (robust=-2) radio image overlaid on HST H$\alpha$ image. The first contours are drawn at 3~$\sigma$, at $\pm$~$45$~$\mu$Jy/beam. The positive contour levels increase by a factor of 1~$\sigma$. The peak brightness is 91.7 $\mu$Jy/beam. The Gaussian restoring beam is $1.6" \times 1.1"$ at PA=$-19^\circ$.}
 \end{figure*}



 

\acknowledgments
The research leading to these results has received funding from the
European Community's Seventh Framework Programme (FP7/2007-2013) under
grant agreement number ITN 215212 "Black Hole Universe".\\

\end{document}